\documentclass[english,12pt]{iopart}
\usepackage[T1]{fontenc}
\usepackage[latin1]{inputenc}
\usepackage{iopams}
\usepackage{graphicx}
\makeatletter
\usepackage{babel}
\usepackage[ps2pdf]{hyperref}
\begin{document}
\title{Unitary theory of laser Carrier-Envelope Phase effects}
\author{V. Roudnev$^1$ and B.D. Esry}
\address{
J.R. Macdonald Laboratory, 
Department of Physics, 
Kansas State University,
Manhattan, KS, 66502 USA}
\address{$^1$ roudnev@phys.ksu.edu} 
\begin{abstract}
We consider a quantum state interacting with a short
intense linearly polarized laser pulse.
Using the two-dimensional time representation and Floquet picture
we establish a straightforward  connection between the laser
carrier-envelope phase (CEP) and the wave function. This connection is
revealed as a unitary transformation in the space of Floquet
components. It allows any CEP effect to be interpreted as an
interference between the components and to put limits on using
the CEP in coherent control.
A 2-level system is used to illustrate the theory. On this example we
demonstrate strong intensity sensitivity of the CEP effects and predict
an effect for pulses much longer than the oscillation period of the
carrier.
\end{abstract}
%\maketitle

\section{Introduction }
Progress in manipulating ultrashort intense laser pulses has made
possible studying laser-matter interactions in qualitatively new
regimes. For instance, very short  pulses having only a few
oscillations of the laser field can be produced \cite{Nature,Nature2}.
In contrast to the more 
conventional case, when a laser pulse is much longer than the carrier
period, the carrier phase of the pulse with respect to the envelope
maximum can become an important parameter for short pulses. 
This phase is called the carrier-envelope (CE) phase (CEP). It has
been demonstrated experimentally, that the CE phase can significantly influence 
ionization of Kr atoms by infrared laser pulses \cite{Nature}.
Recently, similar experiments were performed with Rydberg states of Rb
atoms ionized by a few-cycle 25 MHz pulse \cite{Dutch}, and the spatial 
distribution of the ionized electrons has shown sensitivity to the CEP. Even
potentially stronger effects were predicted theoretically for
dissociation of the HD$^+$ in the laser field \cite{OurPRL} and
experiments are being performed. The sensitivity of high harmonic
generation (HHG) on the laser CEP is also known
\cite{Nature2}. Molecular isomerisation in short intense laser
pulses also provides an interesting example where CEP effects are
important \cite{isomer}. 

However, the CEP effects are, probably, not fully understood theoretically.
Only a few models that go beyond the qualitative picture have been 
discussed. An interesting interpretation of the CEP effects in
ionization as a double-slit interference in the time domain was proposed
by F.Linder et al. \cite{doubleSlit}. This interpretation, however,
does not help to describe results for high energy electrons, or, more
generally, to describe the dependence of the CEP effects on the final
state energy. The goal of this paper is to propose a simple and physically
significant interpretation for the influence of the CE phase on the
interacting wave function. We propose a picture revealing a simple
but exact relationship between the laser phase and the state. 

We use the 2D time (or $(t,t')$) formalism \cite{Moiseev}
together with a Floquet representation of 
the interaction. This approach allows a model-free separation of different
time scales present in a short laser pulse. We shall see how the CEP
can be eliminated from the equations by a simple unitary
transformation. This transformation allows recovery of the CEP
dependence of a final state from the wave function components for one
CE phase only. 

We illustrate our result by demonstrating how the state of a
two-level system (a qubit) can be controlled by a short
pulse, revealing the mechanisms of CEP influence on the state of the
system. We demonstrate and explain dependence of the CEP effect
magnitude on the maximal field of the pulse and the pulse duration. It
is shown that CEP effects can be observed even for pulses that are
much longer than one oscillation period of the carrier.

\section{Theory}
We base our approach on investigating the time-dependent Schr\"odinger equation
\begin{equation}
  i\frac{\partial}{\partial t}\Psi=[H_0+V(t)]\Psi \ .
  \label{TDSE}
\end{equation}
Here $\Psi$ is the wave function, $H_0$ is the Hamiltonian of the system
without laser field, and $V(t)$ stands for 
the laser-matter interaction potential.
We shall consider laser-matter interactions of the following form
\begin{equation}
  V(t)=-{\bf E}(t)\cdot{\bf d}\cos(\omega t+\varphi) \ \ .
  \label{LaserField}
\end{equation}
Here ${\bf E}(t)$ is the envelope of the laser pulse field, {\bf d}
 is the dipole interaction operator, $\omega$ is the laser carrier
frequency and $\varphi$ is the carrier-envelope phase. The latter
can be a very important parameter especially for sufficiently short
pulses, and it is the main parameter studied in this work. 

In what follows, we give a brief description of the 2D time formalism 
\cite{Moiseev} and introduce a two-dimensional
time representation for a system in a pulsed laser field. We shall
demonstrate how this representation allows eliminating of the CEP
$\varphi$ from the evolution equation and construction of a
CEP-independent solution. The CEP dependent
solution of the initial equation (\ref{LaserField}) 
can be recovered from the CEP-independent solution with a unitary
transform. 

\subsection{The 2D time formalism for a system in a periodic external field}

The formalism of two-dimensional time is very useful to treat time-dependent
systems that show both periodic and non-periodic behavior, such as
atoms and molecules in a field of a laser pulse. One of the advantages
of this formalism is that it allows separation of periodic and non-periodic
dynamics without resorting to adiabatic expansions that might converge
slowly for nonadiabatic systems. 

In the 2D-time representation one introduces a second time
$s$, such that the envelope and the periodic factors in the laser-matter
interaction (\ref{LaserField}) depend on different time coordinates
\begin{equation}
\begin{array}{l}
  V(s,t)=-({\bf E}(s),{\bf d})\cos(\omega t+\varphi) \\
  H(s,t)=H_0+V(s,t) \ \ \ .
\end{array}
\label{2DTHamiltonian}
\end{equation}
One also introduces a second time dynamics
\begin{equation}
(i\frac{\partial}{\partial s}+i\frac{\partial}{\partial t})\Psi(s,t)
=
(H_{0}+V(s,t;\varphi))\Psi(s,t)\ \ \ .
\label{2TDSE}
\end{equation}
It is not difficult to see that if $\Psi(s,t)$ satisfies (\ref{2TDSE}),
the solution of (\ref{TDSE}) can be written as $\Psi(t)=\Psi(s,t)\mid_{s=t}$.
Indeed, restricting the solution to the {}``diagonal'' time $s=t$
and substituting it to the left-hand side of the equation (\ref{TDSE})
we get the left-hand side of the equation (\ref{2TDSE})
\[
i\frac{\partial}{\partial t}\Psi(t,t)=(i\frac{\partial}{\partial s}
                                     +i\frac{\partial}{\partial t})\Psi(s,t)\mid_{s=t}\ \ \ .
\]
The right-hand sides of the equations (\ref{TDSE}) and (\ref{2TDSE})
are identical at $s=t$. Thus, once the 2D-time equation is solved,
one has a solution of the original equation (\ref{TDSE}) as well.
More detailed discussion of the 2D-time formalism can be found at
the original paper of Peskin and Moiseev \cite{Moiseev}. 
%%%%%%%%%%%%%%%%%%%%%%%%%%%%%%%%%%%%%%%%%%%%%%%%%%%%%%%%%%%%
\subsection{Floquet representation for a finite-time pulse }
%%%%%%%%%%%%%%%%%%%%%%%%%%%%%%%%%%%%%%%%%%%%%%%%%%%%%%%%%%%%
If the laser pulse duration is not considerably shorter than the oscillation
period $2\pi/\omega$, expanding the wave function $\Psi(s,t)$
into a Fourier series in $t$ can be a reasonable way of solving 
equation (\ref{2TDSE}), even for pulse duration comparable with the
oscillation period \cite{ShortPulseFloquet}. In fact, this approach
holds as far as the description of the laser pulse itself in terms of carrier
and envelope is valid \cite{RevModPhys}. In this paper we discuss
pulses longer than one oscillation period, and, thus, our approach to solving
equation (\ref{2TDSE}) is justified. 

We start constructing the Floquet representation
\cite{Floquet1,Floquet2} for the laser pulse
from expanding the wave function 
\begin{equation}
  \Psi(s,t)=\sum_{n=-\infty}^{\infty}e^{in\omega t}\phi_{n}(s)\ \ \ .
\label{Fourier}
\end{equation}
We shall call the coefficients $\phi_{n}(s)$ $n$-photon emission ($n>0$)
and absorption ($n<0$) amplitudes. 

Bringing the time derivative $i\frac{\partial}{\partial t}$ to the right hand side
of equation (\ref{2TDSE}) and substituting the representation
(\ref{Fourier}) we come up with the following infinite system of
equations
\begin{equation}
\begin{array}{c}
i\frac{\partial}{\partial s}\phi_{n}(s)=
  \begin{array}[t]{r}
    \frac{1}{2}e^{i\varphi}({\bf E}(s),{\bf d})\  \phi_{n-1}(s)\\
     +(H_{0}+n\omega)\  \phi_{n}(s) \\
     +\frac{1}{2}e^{-i\varphi}({\bf E}(s),{\bf d})\  \phi_{n+1}(s)
  \end{array}\\
  n=-\infty,\ldots,-1,0,1,\ldots,\infty
\end{array}\ \ \ .
\label{TDFloquet}
\end{equation}
Starting at time $s=s_{-}$ from the initial state 
\[
\begin{array}{lr}
\phi_{0}(s_{-})=\psi_{0}\\
\phi_{n}(s_{-})=0 & n\neq0
\end{array}
\]
 and propagating the amplitudes according to (\ref{TDFloquet}) sufficiently
long, we can investigate different laser-induced processes, such as
dissociation, ionization, (de)excitation etc.

It is convenient to introduce a vector of n-photon amplitudes 
${\bf \Phi}(s;\varphi)$
and a Floquet Hamiltonian
\[{\bf H}(s;\varphi)\equiv{\bf H}_0+{\bf V}(s;\varphi)\] 
where ${\bf H}_0$ stands for the diagonal and ${\bf V}(s;\varphi)$ for
the off-diagonal terms in the equation (\ref{TDFloquet}).
The equation (\ref{TDFloquet}) can be written as 
\[
  i\frac{\partial}{\partial s}{\bf \Phi}(s;\varphi)={\bf H}(s;\varphi){\bf \Phi}(s;\varphi)\ \ \ .
\]
 In this notation we emphasize that the Floquet Hamiltonian and the
solution both depend on the carrier-envelope phase $\varphi$. This
dependence, however, is parametric, and it can be eliminated.

%%%%%%%%%%%%%%%%%%%%%%%%%%%%%%%%%%%%%%%
\subsection{Unitary equivalence of the CEP}
%%%%%%%%%%%%%%%%%%%%%%%%%%%%%%%%%%%%%%%
The obvious advantage of equation (\ref{TDFloquet}) for  studying the CEP effect
is that the phase dependence enters the equation linearly. As we
shall see, it allows exclusion of the CE phase from
the Floquet Hamiltonian by a simple unitary transformation. 

Let us introduce the following operator ${\bf U}(\varphi)$ acting
in the n-photon amplitude space
\[
[{\bf U}(\varphi)]_{mn}=\delta_{mn}e^{-in\varphi}\ \ \ .
\]
Now consider the Floquet Hamiltonian ${\bf H}(0)$corresponding to
$\varphi=0$. It is easy to verify that the following 
relation holds:
\[
{\bf H}(\varphi)={\bf U^\dagger}(\varphi){\bf H}(0){\bf U}(\varphi)\ \ \ 
\]
and to rewrite the evolution equation as 
\[
i\partial_{s}{\bf U}(\varphi){\bf\phi}(s;\varphi)={\bf H}(0){\bf U}(\varphi){\bf\phi}(s;\varphi)\ \ \ .
\]
This equation helps to establish the main result of this paper, which is the unitary
equivalence of solutions corresponding to different carrier-envelope
phases
\begin{equation}
  {\bf \Phi}(s;\varphi)={\bf U^{\dagger}}(\varphi){\bf \Phi}(s;0)\ \ \ \ .
\label{theMain}
\end{equation}
This relationship allows any CEP effect to be interpreted as interference
of different n-photon channels. It can be clearly
seen if we recall the expression for the wave function in physical
time
\begin{equation}
  \Psi(t;\varphi)=\sum_{n=-\infty}^{\infty}e^{in\varphi}\tilde{\phi}_{n}(t;0)\ \ ,
  \label{PhaseImprint}
\end{equation}
where we have introduced the final state components of the wave
function
$\tilde{\phi}_{n}(t;0)=e^{in\omega t}\phi_{n}(t;0)$. 
After the laser pulse is off at the moment $t=t_0$, the components
$\tilde{\phi}_{n}$ depend on time uniformly due to the internal dynamics
of the system only, i.e. ${\bf \phi}(t)=e^{-i{\bf H}_0(t-t_0)}{\bf \phi}(t_0)$. 
Depending on the laser CEP, the wave function components
$\tilde{\phi}_{n}$ gain phases proportional to the corresponding net number of
exchanged photons $n$ and the CEP $\varphi$.

It is important to note that all the given results are obtained under
rather general assumptions of dipole laser-matter interaction and a stable
shape of the laser pulse. We did not assume anything specific
for any particular physical system. That means that any CEP effect
can be considered as interference of several n-photon channels.

\subsection{CEP effect observation}
Before discussing application of the formula (\ref{PhaseImprint}) to
particular model systems, we say a few words on observing the phase
effects in general. 

Let $\hat{O}$ be an observable of interest. Using the representation 
(\ref{PhaseImprint}) at large times we get an explicit CEP dependence of the
mean value of $\hat{O}$ through the n-photon components at zero phase
\[
  \langle\hat{O}\rangle=\sum_{k,n=-\infty}^{\infty} e^{i (k-n)\varphi} 
                 \langle \phi_n(t_0;0)|\hat{O}(t)|\phi_k(t_0;0)\rangle
   \ \ \ ,
\]
where 
$\hat{O}(t)=e^{i{\bf H}_0(t-t0)}\hat{O}e^{-i{\bf H}_0(t-t0)}$.
Rearranging the terms in the series and using $\hat{O}^\dagger=\hat{O}$ we
get a Fourier series for the $\langle\hat{O}\rangle$ CEP dependence 
\begin{equation}
  \langle\hat{O}\rangle(\varphi)=\frac{\alpha_0}{2}
           +\sum_{k=1}^{\infty} 
	       ({\rm Re}\alpha_k \cos k \varphi+
	       {\rm Im}\alpha_k\sin k \varphi) \ \ \ 
\label{phiDep}
\end{equation}
with the coefficients $\alpha_k$ defined as

\[
  \alpha_k=2\sum_{n=-\infty}^{\infty} \langle \phi_{n-k}(t_0;0)|\hat{O}(t)|\phi_n(t_0;0)\rangle \ \ \ .
\]

It is useful to introduce a measure of CEP effect observability. This
measure can be chosen as a norm of the $\varphi$-dependent part in 
equation (\ref{phiDep}) 
\begin{equation}
  \sigma = (\sum_{k=1}^{\infty}|\alpha_k|^{2})^{\frac{1}{2}} \ \ \ .
  \label{observability}
\end{equation}
We shall refer to this quantity as absolute CEP amplitude of the observable
$\hat{O}$, since it indicates how much $\langle\hat{O}\rangle$ can deviate from
its mean value $\alpha_0$ when varying the CEP. It is worth mentioning
that $\sigma$ is directly connected to the mean square deviation of
$\langle\hat{O}\rangle$ from its CEP-averaged value, namely:
\[
 [\int_{0}^{2\pi}(\langle\hat{O}\rangle(\varphi)-\frac{\alpha_0}{2})^2d\varphi]^{\frac{1}{2}}
  =\sqrt{\pi} \sigma \ \ \ .
\]
It is also useful to have a CEP amplitude weighted with the mean value
$\alpha_0$ and experimental sensitivity $\delta$
\begin{equation}
   \sigma_W=\frac{\sigma}{\frac{\alpha_0}{2}+\delta} \ \ \ .
   \label{aweighted}
\end{equation}

\section{Numerical illustrations}

In the following section we give two types of demonstrations: quantitative and
qualitative. 

Quantitatively, we shall check the agreement of equations
(\ref{theMain}) and (\ref{PhaseImprint}) 
with CEP results calculated independently. For that type of demonstration we have to
calculate the n-photon amplitudes either from a direct solution of
equation (\ref{2TDSE}) or by extracting them from an independently calculated
final state wave function. Knowledge of the amplitudes, however,
is rather expensive. To predict the CEP response of any particular
physical system, we have to know not
only the population probabilities of the Floquet states, but
their phases as well. Usually, this information can be obtained only from a
numerical solution of the TDSE, and we shall explain how expression 
(\ref{PhaseImprint}) can be used to reduce the amount of numerical 
calculations.

Qualitatively, we shall discuss the physical conditions needed to
observe CEP effects experimentally. Such conditions are equivalent
to the existence of several interfering components. In fact, as equations 
(\ref{theMain},\ref{PhaseImprint})
suggest, CEP effects exist if and only if several n-photon components
contribute to the same physical state. We shall see how this condition
is realized in a simple 2-level model. On this example we shall 
discuss intensity and pulse duration dependence of the CEP effects.
We should mention, however, that not all the physical systems conventionally
treated in a 2-level approximation are suitable for clear CEP effect
demonstrations. For instance, choosing an experimental realization,
one has to fulfill the condition that the two states used in the
demonstration must be well separated from other eigenstates of the
system, because the important 
transitions have essentially nonresonant multiphoton character.
Such systems can be realized, for example, as double quantum dots
\cite{QDot1,QDot2} or as ionic hyperfine qubits \cite{QIP}.
Although the 2-level model cannot describe 
processes of ionization or molecular dissociation realistically, 
it still gives a reasonable qualitative description of the CEP effect
observability. More detailed study of the CEP effects involving
continuum states is a subject for another investigation.

In all the examples we shall use a Gaussian shape of the laser
pulse
\[
E(t)=E_{0}e^{-(\frac{t}{\tau})^{2}}\ \ ,
\]
\[
V(t)=-d\cdot E(t)\equiv V_{0}e^{-(\frac{t}{\tau})^{2}}\ \ ,
\]
where $E_0$ is the peak field, $\sqrt{2\ln 2} \tau$ is the
intensity FWHM pulse duration and $V_0\equiv -d\cdot E(T)$ is the peak
interaction energy. Since we are particularly interested in molecular
systems, we choose energy scales in the typical energy range of molecular
vibrational states and fix the laser carrier frequency to 0.058~a.u.,
what corresponds to a standard 790~nm Ti:Sapphire laser. 

\subsection{Excitations in a 2-state model.}
We start from a simplest example of an excitation in a 2-state system.
Let $|a\rangle$ and $|b\rangle$ be
the two eigenstates with energies $E_{a}$ and $E_{b}$. Without loss
of generality we can set the first energy to zero, $E_{a}\equiv0$,
$E_{b}\equiv E_{a}+\Delta =\Delta $, and consider $|a\rangle$ the initial
state. The wave function in this case takes form 
\[
  \psi=a(t;\varphi)|a\rangle + b(t;\varphi) |b\rangle \ \ \ .
\]
Suppose the coupling to the laser field
between the two states is proportional to the laser field, such that
the corresponding time-dependent Scr\"odinger equation reads
\begin{equation}
i\frac{\partial}{\partial t}
     \left(\begin{array}{c}
            a(t) \\
	    b(t)
           \end{array} \right)
=\left(\begin{array}{cc}
           0 & V(t)\cos(\omega t+\varphi)\\
     V(t)\cos(\omega t+\varphi) & \Delta \end{array}
 \right) 
 \left(\begin{array}{c}
            a(t) \\
	    b(t)
           \end{array} \right) \ .
\label{TDSE2L}
\end{equation}
After introducing the coupling energy $V(t)=-d\cdot E(t)$ and keeping only
states coupled to the initial state $|a\rangle$ the 
Floquet Hamiltonian ${\bf H}$ for $\varphi=0$ in (\ref{TDFloquet}) can be written
as
\[
{\bf H}=
\left(
     \begin{array}{ccccccc}
       \ldots & \frac{V(s)}{2} & 0 & 0 & 0 & 0 & 0\\
         \frac{V(s)}{2} & 0-2\omega & \frac{V(s)}{2} & 0 & 0 & 0 & 0\\
          0 & \frac{V(s)}{2} & \Delta-\omega & \frac{V(s)}{2} & 0 & 0 & 0\\
                 0 & 0 & \frac{V(s)}{2} & 0 & \frac{V(s)}{2} & 0 & 0\\
          0 & 0 & 0 & \frac{V(s)}{2} & \Delta+\omega & \frac{V(s)}{2} & 0\\
         0 & 0 & 0 & 0 & \frac{V(s)}{2} & 0+2\omega & \frac{V(s)}{2}\\
                 0 & 0 & 0 & 0 & 0 & \frac{V(s)}{2} & \ldots
     \end{array}
\right) \ \ \ .
\]
In this representation even Floquet amplitudes of 
correspond to the ground state $|a\rangle$  and odd ones to the
excited state $|b\rangle$, i.e. 
\[
\begin{array}{l}
  |\phi_{2n}\rangle\equiv \tilde{a}_{2n}(t) e^{-2n\omega t}|a\rangle \ \ , \\
  |\phi_{2n+1}\rangle\equiv \tilde{b}_{2n+1}(t) e^{-(2n+1)\omega t}|b\rangle  \ \ , \\
  n=-\infty,\ldots,\infty   \ \ \ .
\end{array}
\]  

\begin{figure}[tbp]
\begin{center}
\includegraphics[scale=0.6,clip=true]{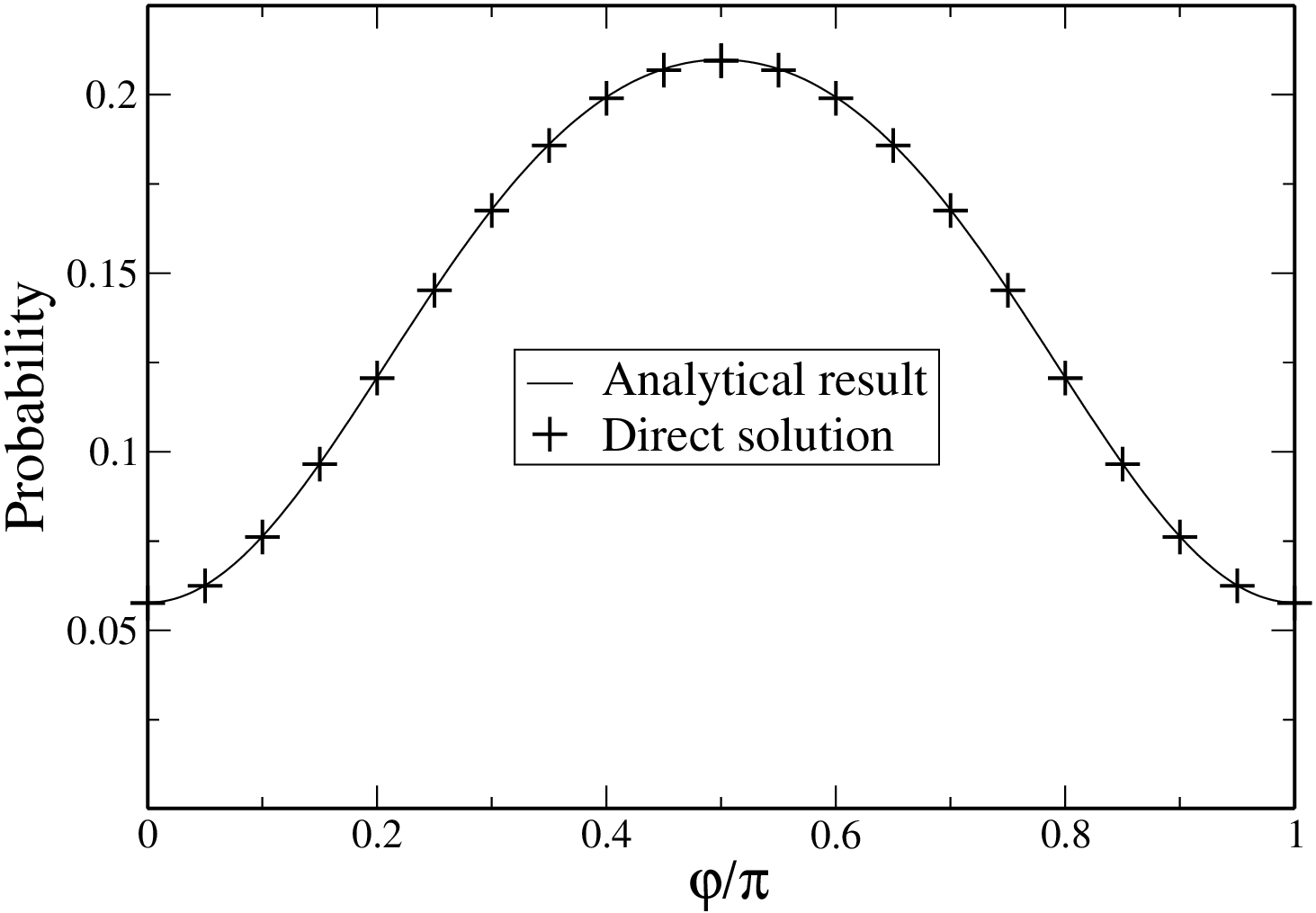}
\end{center}
\caption{\label{FigCEPProb}2-level system photoexcitation probability
as a function of the CEP for the energy gap $\Delta=0.066$~a.u., 
laser frequency $\omega=0.058$~a.u., peak field $V_0=0.1194$~a.u.,
pulse duration $\tau=248$~a.u.. The dots are obtained from the direct
solution of equation (\ref{TDSE2L}), the line is constructed from
the final state n-photon amplitudes for $\varphi=0$.}
\end{figure}

Let us demonstrate first how the CEP reveals itself in the excitation
probability. At first, we calculate the Floquet amplitudes for
$\varphi=0$ numerically. In this example we use the energy gap
$\Delta=0.066$~a.u., which slightly exceeds the photon energy, and the
dipole coupling is chosen as $d=-1$~a.u.  
The pulse intensity is chosen as $5\times 10^{14}$~W/cm$^2$, which
corresponds to the peak interaction energy $V_0=0.1194$~a.u., and 
the pulse duration is $\tau=248$~a.u., which is about 7~fs intensity FWHM. 
The calculations were performed with 30 Floquet blocks, what
guarantees accurate results even for intensities higher than
$10^{15}$~W/cm$^2$.
Let us look for the excitation probability $P_{ex}$
as the observable.  
We choose the final propagation time $t_0=4 \tau
=992$~a.u., when the field is negligible.
According to equation (\ref{PhaseImprint}), the final state wave
function is expressed in terms of the Floquet amplitudes as
\begin{equation}
\begin{array}{rcl}
\Psi(t;\varphi) &=&\left(\sum_{n=-\infty}^{\infty} e^{i2n\varphi}
\tilde{a}_{2n}(t_0)\right) |a\rangle \\
   &+& \left(\sum_{n=-\infty}^{\infty} e^{i(2n+1)\varphi} \tilde{b}_{2n+1}(t_0)\right)
   e^{-\Delta (t-t_0)} |b\rangle \ \ .
\end{array}
   \label{PhaseImprint2L}
\end{equation}
Thus, in order to calculate the excitation probability
we have to evaluate the odd n-photon components of the
final state
\[
%\begin{array}{rcl}
  P_{ex}=|\langle b|\Psi\rangle|^2 %\\
        =  |\sum_{n=-\infty}^{\infty} e^{i(2n+1)\varphi}\tilde{b}_{2n+1}(t_0)|^2 \\
%\end{array}
\]
The following final state amplitudes will
contribute to the excited state of the system at the final time
$\tilde{b}_{1}(4 \tau;0)= 0.0092 -0.0202 i  $, 
$\tilde{b}_{-1}(4 \tau;0)=  0.1445 -0.3178 i$, 
$\tilde{b}_{-3}(4 \tau;0)= -0.0543+0.1193 i$. All other amplitudes are
negligibly small. According to equation (\ref{PhaseImprint2L}), the excited
state component CEP dependence reads 
\[ 
b(t;\varphi)|b\rangle=(\tilde{b}_{1}e^{i\varphi}
   +\tilde{b}_{-1}e^{-i\varphi}
   +\tilde{b}_{-3}e^{-3i\varphi}) e^{-i\Delta (t-t_0)}|b\rangle \ \ \ .
\]
This results in the following explicit expression for the excitation
probability:
\[
  P_{ex}=0.1395 - 0.07602 \cos 2\varphi - 0.00582 \cos 4\varphi \ \ .
\]
This line is shown in the Fig.~\ref{FigCEPProb} together with results
of 
direct solution of equation (\ref{TDSE2L}). Since no
approximations were made, besides cutting off the series
(\ref{PhaseImprint2L}), the perfect agreement is not surprising. 

Giving this example, we calculated the Floquet amplitudes by direct
solution of the system of equations (\ref{TDFloquet}). In practical
applications, however, this approach is not effective: usually,
solving a small system $n$ times is simpler than solving an $n$ times 
bigger system of equations once. In the case of the equations
(\ref{TDFloquet}) the situation is even worse. Even if we have only 3
components contributing to the final state, as in our example, one
needs more than 5  Floquet blocks kept in the equations to reproduce the
correct dynamics even for moderate peak fields. Because of that, recovering the amplitudes by
subsequently solving the original Schr\"odinger equations directly for
several CE phases and fitting the expression (\ref{PhaseImprint}) to
the wave function should be a much more effective procedure for
realistic calculations.  

After giving this numerical example, we are ready to discuss some
qualitative properties of the CEP effects.
Are there any conditions for the peak interaction energy $V_{0}$ 
and the pulse duration $\tau$ that limit an observability of CEP
effects? 

\begin{figure}[tbp]
\begin{center}
\includegraphics[scale=0.6,clip=true]{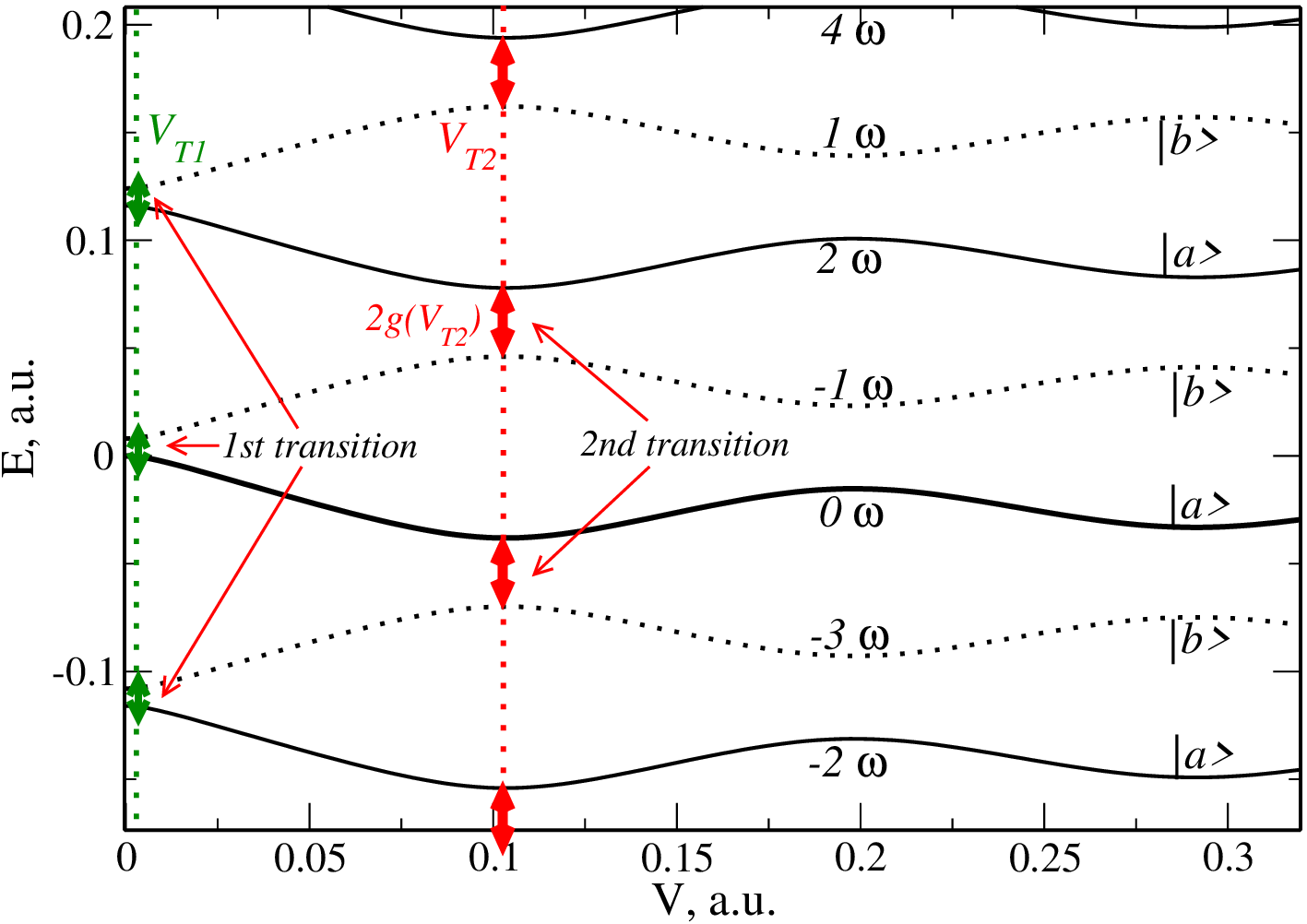}
\end{center}
\caption{\label{cap:2-level-system-Floquet}2-level system Floquet Hamiltonian
eigenstates as a function of the laser-matter interaction energy
$V=-E\cdot d$.
The model parameters are $\omega=0.058$a.u., $\Delta=0.066$a.u. Thick lines
correspond to the ground state and dotted lines to the excited state.
}
\end{figure}

\begin{figure}[tbp]
\begin{center}
\includegraphics[scale=0.6,clip=true]{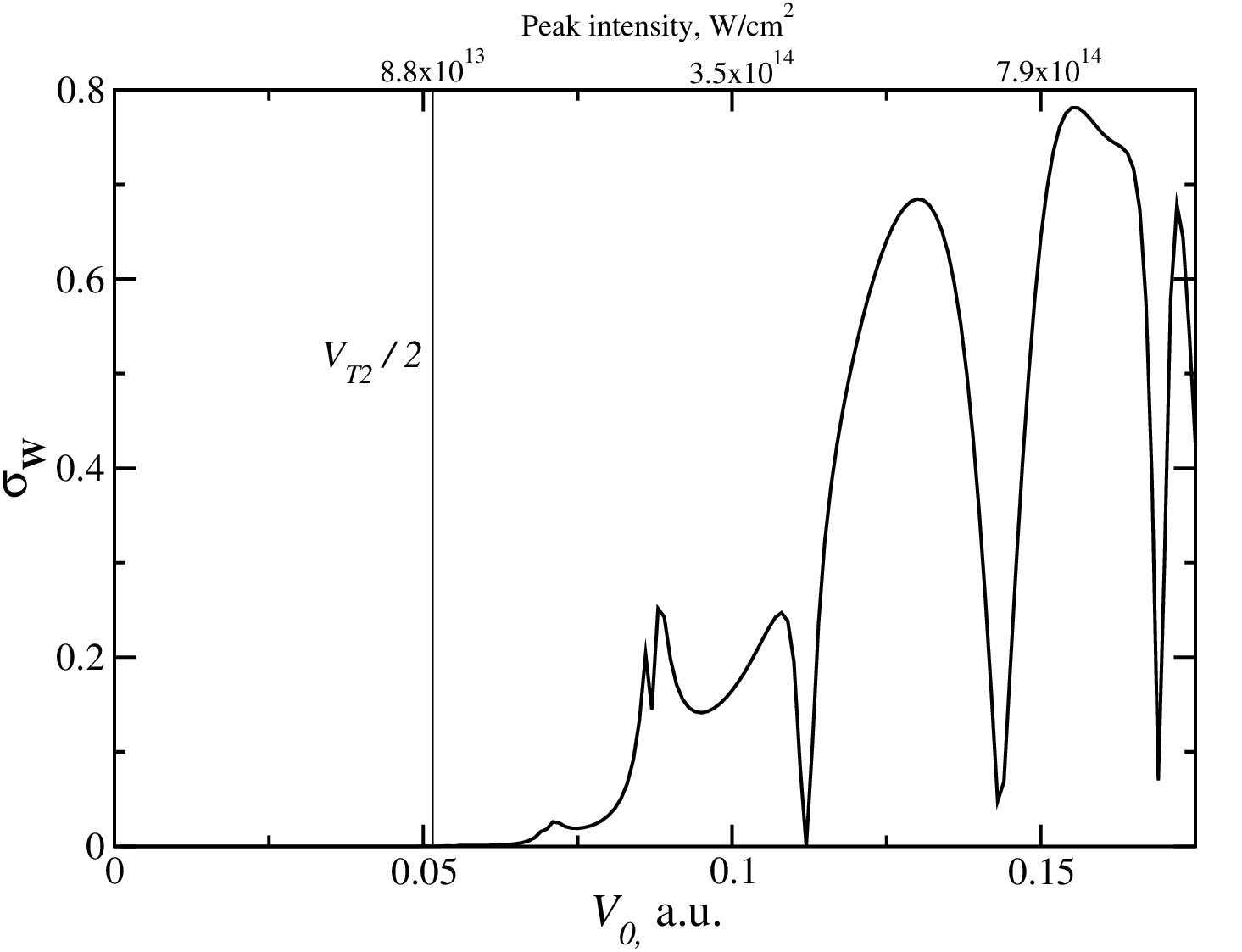}
\end{center}
\caption{The weighted excitation probability CEP amplitude $\sigma_W$
(\ref{aweighted})
as a function of the peak interaction energy. 
No phase dependence
below $V_0=0.05$~a.u. is observed. Calculations were performed with 30
Floquet blocks, the sensitivity parameter $\delta$ in (\ref{aweighted})
is set to $0.01$.
\label{cap:CEP-dependence-contrast}
}
\end{figure}

Let us discuss the intensity dependence first. In order to have
interference, one has to make sure that there  
are several $n$-photon components contributing to the states with the
same final state energy. It cannot be achieved with one-photon transitions
only, so there must be a minimal intensity that allows a clear
observation of the CEP effect. To understand the lower intensity
limit, consider the eigenvalues 
of ${\bf H}$ shown in Fig.~\ref{cap:2-level-system-Floquet}. 
On the leading edge of the laser pulse the interaction energy grows
up, this correponds to moving from the left to the right in
Fig.~\ref{cap:2-level-system-Floquet}. After the peak, which defines
the rightmost point in Fig.~\ref{cap:2-level-system-Floquet}, the
system goes from the right to the left during the trailing edge
of the pulse. 
As one can see from tracking the eigenvalues of the
dressed system, the two crossings that we need to populate the
state $|b\rangle$ via 1- and 3-photon transitions correspond to the interaction
energies $V_{T1}\approx 0$~a.u. and $V_{T2}\approx 0.1$~a.u.. This gives us
a limitation on the laser peak intensity: no interference can happen if the maximal field-matter
interaction energy does not approach the second crossing.
We can safely say that in our example no CEP effect is expected if the peak
interaction energy is smaller than $0.05$~a.u.. This observation is
illustrated in Fig.~\ref{cap:CEP-dependence-contrast}, where
we have plotted the weighted  CEP effect amplitude (\ref{aweighted}).
As we expected, we see an essential growth of the CEP effect
contrast only above $0.05$~a.u., when the peak field starts approaching
the 3-photon crossing. 

The big maximal intensity alone, however, is not enough to observe
the effect: the field should change fast enough when passing through
both crossings, otherwise transition happens adiabatically, and
no population transfer occurs. The proper timing conditions must be
satisfied. This leads us to the question of the pulse duration
dependence of the CEP effects.

\begin{figure}[tbp]
\begin{center}
\includegraphics[scale=0.6,clip=true]{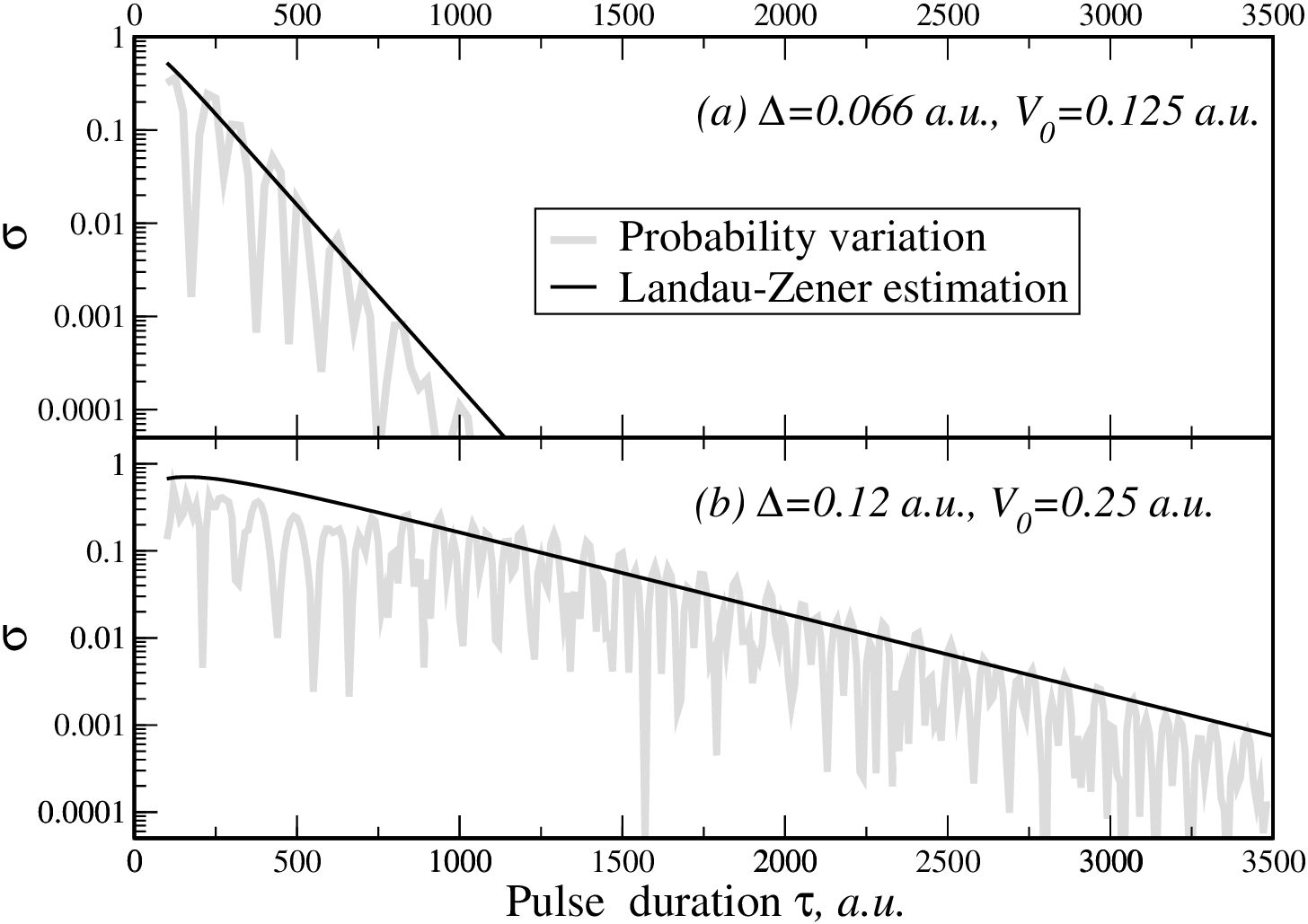}
\end{center}
\caption{\label{Fig:Tau}CEP variation of the excitation probability as a
function of the pulse duration. The black lines correspond to the
Landau-Zener estimation of the effect via the transition probability
at the second crossing. a) $\Delta=0.066$~a.u. b) $\Delta=0.12$~a.u.
The peak interaction energy is taken about 30\% higher than the 
second transition energy.}
\end{figure}

As we have seen, nonadiabatic second transition (see
Fig.~\ref{cap:2-level-system-Floquet}) is a necessary condition for
the CEP observation. We can use 
the Landau-Zener model to qualify the presence of the CEP effects by
estimating the second transition probability as a function of the
pulse duration. Let $V_{T2}$ be
the position of the second crossing. 
For the energy gap of our example ($\Delta=0.066$~a.u.) the level splitting is
about $g(V_{T2})=0.016$~a.u.\footnote{For small energy gaps
compared to the photon energy we can use a Bessel approximation to
the spectrum of the Floquet Hamiltonian. This allows us to estimate 
the splitting of the Floquet eigenstates at the crossing point as 
$
  g(V_{T2})\approx\frac{1}{2}|(\omega-D\frac{2}{\pi} \sqrt{\frac{2}{5}})|
   \approx \frac{1}{2}|(\omega-0.40 D)|
$.}
The growth rate of the interaction energy when crossing the second transition
region
\[
\frac{dV}{dt}=2 \frac{V_{T2}}{\tau}\sqrt{\ln \frac{ V_0}{V_{T2}}}
\]
 plays the role of velocity in the Landau-Zener formula. 
Now we are ready to estimate the second transition
probability as
\begin{equation}
\begin{array}{rcl}
P_2&=&e^{-2 \pi \delta}(1-e^{-2 \pi \delta}) \\
\delta&=&\frac{g(V_{T2})^2 \tau}{2 V_{T2}\sqrt{\ln (V_0/V_{T2})}} \\
\end{array}
\label{LZtime}
\end{equation}
If no higher order transitions happen, the amplitude of the CEP
effect should be proportional to the square root of this probability,
since it is proportional to the amplitude rather than the probability
of the m-photon state population.
This square root of the probability is shown in Fig.~\ref{Fig:Tau}a together 
with the amplitude of the excitation probability CEP dependence. 
It is clear that probability (\ref{LZtime}) should have a
substantial value to 
make possible an observation of the CEP effects. Equation
(\ref{LZtime}) shows that CEP effects should disappear
exponentially with growing pulse length. 

It is important to mention that 
the pulse duration needed to observe the CEP effects is a property of the
laser-matter interaction rather than the laser pulse alone. In the
two-level model, as equation (\ref{LZtime}) suggests,
it is defined by level splitting at the second crossing point and the position
of the second crossing. For large energy gaps it is easy to
obtain smaller multiphoton splitting. This allows one to predict the CEP
effects even for pulses that are substantially longer than one
oscillation period, as demonstrated in Fig.~\ref{Fig:Tau}b.
There we show the CEP observability $\sigma$ together with its Landau-Zener
estimation. Even for pulses longer than $\tau=1700$~a.u., what is
about 30 periods in the field FWHM, we still can see a variation of
the probability with the CEP about 10\%. 
We must mention, however, that experimental observability of the long-pulse
effect can be limited by the stability of the pulse shape, which might
be difficult to keep at a time scale smaller than one oscillation
period for long pulses.

\section{Summary}
We have investigated a quantum state experiencing a laser pulse of a
stable envelope shape and varying CEP of the oscillatory part. We
have shown how the CEP can be excluded from evolution equations and
how the CEP dependence of the final state can be recovered from
CEP-independent results. On the example of a 2-level system we have
demonstrated a critical intensity and pulse duration dependence of CEP effects. 
In contrast to the common conception that CEP effects can be
expected only when the pulse duration is nearly as short as the laser oscillation period, 
we have demonstrated that CEP effects can exist even for
pulses much longer than that. The pulse duration that allows the CEP
effect observation critically depends on the properties of the system
interacting with a laser pulse. In the 2-level system, long-pulse
CEP effects can be observed in essentially above-threshold excitation
regimes, when the energy gap is considerably larger than the photon
energy.

The approach which is suggested in this paper is rather general, and it
would be interesting to study more complex systems from this point of
view. For instance, it would be interesting to study how $n$-photon
component interference affects high-harmonic generation, ionization
and dissociation in different systems. There are also many
physical and mathematical questions to be addressed, such as gauge-invariant
formulation of the theory and checking the classical limits of the
interference effect.

\section*{Acknowledgments}
This work was supported by the Chemical Sciences, Geosciences,
and Biosciences Division, Office of Basic Energy Sciences, Office of Science, U.S.
Department of Energy. Authors wish to thank Prof. Ben-Itzhak and
Prof. Cocke for stimulating discussions, and Carol Regehr for reading
and editing the draft.

\end{document}